\documentclass[prab,twocolumn,amsmath,amssymb]{revtex4-2} 

\usepackage[colorlinks=true,bookmarks=false]{hyperref}
\usepackage{graphicx}
\usepackage{dcolumn}
\usepackage{bm}

\begin{document}

\title{LCODE: Quasistatic code for simulating long-term evolution of three-dimensional plasma wakefields}

\author{I.Yu.~Kargapolov, N.V. Okhotnikov, I.A. Shalimova, A.P. Sosedkin, and K.V. Lotov}
\affiliation{Novosibirsk State University, Novosibirsk 630090, Russia}
\date{\today}

\begin{abstract}
A recently developed three-dimensional version of the quasistatic code LCODE has a novel feature that enables high-accuracy simulations of the long-term evolution of waves in plasma wakefield accelerators. 
Equations of plasma particle motion are modified to suppress clustering and numerical heating of macroparticles, which otherwise occur because the Debye length is not resolved by the numerical grid. 
The previously observed effects of premature wake chaotization and wavebreaking disappear with the modified equations.
\end{abstract}

\maketitle

\section{Introduction}
\label{s1:intro}

Plasma is now being actively studied as a medium capable of accelerating particles to high energies at short distances. 
Among various plasma-based acceleration techniques, wakefield acceleration \cite{PoP27-070602, PPCF61-104001, RMP81-1229} provides the highest particle energies \cite{PRL122-084801,Nat.445-741} because the accelerating wave is produced by an object (called a driver) traveling at approximately the speed of light, allowing in-phase acceleration over long distances.

The temporal and spatial scales of accelerating structures and accelerated beams in plasmas are typically very small, tens of femtoseconds and microns. 
Processes at these scales are difficult to characterize experimentally with high resolution \cite{RMP90-035002}, so numerical simulations always play a key role in this field \cite{PoP27-070602, RAST9-165}.
However, direct modeling of plasma wakefield acceleration based on first-principles equations is often too computationally intensive \cite{NIMA-909-476, PPCF61-104004} due to the huge scale difference between structures to be resolved and propagation distances \cite{NIMA-410-461}.
For this reason, various simplified models are widely used \cite{RAST9-165}, one of which is the so-called quasistatic approximation (QSA) \cite{PRL64-2011, PoP4-217} implemented in several codes \cite{PoP22-023103, PRE49-4407, PoP5-785, NIMA-829-350, JCP250-165, CPC261-107784, PPCF56-084012, CPC278-108421, CAS2014-181, PRAB21-071301, AIP1812-050005, PRAB25-104603}.

\begin{figure}[b]
\includegraphics{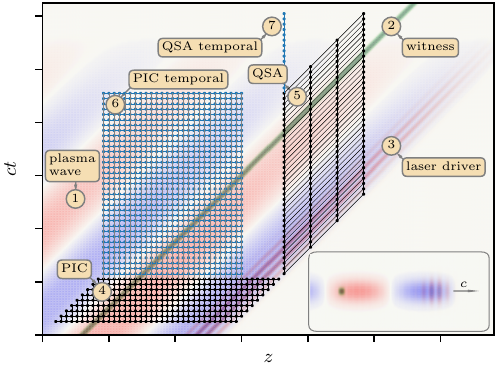}
\caption{Objects simulated in the context of plasma wakefield acceleration (1-3) and computational grids used for simulations of beam dynamics (4,5) and wave evolution (6,7) in PIC (4,6) and QSA (5,7) codes. The inset schematically shows driver, witness, and plasma wave with the same coloring.}\label{fig1-qsa}
\end{figure}

The QSA is based on the change of variables from time $t$ and longitudinal coordinate $z$ to
\begin{equation}\label{e1}
    s = z, \qquad \xi = z-ct,
\end{equation}
where $c$ is the speed of light. 
Most features simulated in the context of plasma wakefield acceleration have a typical 45-degree-elongated shape on the $(z,ct)$ plane (Fig.\,\ref{fig1-qsa}). 
The substitution (\ref{e1}) allows to shear the simulation grid as shown in Fig.\,\ref{fig1-qsa}, make a long grid step in $s$, and thereby increase the computation speed compared to the particle-in-cell (PIC) method \cite{PPR49-229}. 
The speed gain is roughly the ratio of the two scales of quantities' variation, along the line $\xi = \text{const}$ and along the line $t = \text{const}$, and can exceed four orders of magnitude for high energy beams \cite{PPCF62-125023}.

The simulation window in QSA moves at the speed of light, so nothing can propagate forward in it. 
The beams (laser or particle driver and accelerated particles) move backwards slowly in the window, while the plasma particles pass the window at approximately the speed of light. 
Therefore, it is convenient to treat beams and plasma differently. 
For the particle beams, $\xi$ is a space-like variable, and locations $\vec{r} = (x,y,\xi)$ and momenta $\vec{p} = (p_x, p_y, p_z)$ of beam particles are calculated as functions of $s$. 
Advancing $s$ is done similarly to advancing time in PIC codes.
Laser drivers, if any, are treated similarly to particle beams in the sense that the laser field (characterized by the envelope \cite{PoP4-217}) is computed in the three-dimensional (3d) space $(x,y,\xi)$ and advanced in $s$ as if $s$ is the propagation time.
For the plasma particles, $\xi$ is a time-like variable, and two-dimensional (2d) coordinates $\textbf{r}_\perp = (x,y)$ and 3d momenta $\vec{p}$ are calculated as functions of $\xi$ at each $s$ (we highlight 2d vectors in bold).
Advancing $\xi$ is similar to advancing time in PIC codes.
It is additionally assumed that fields and parameters of plasma particles depend on $s$ only through beam properties, which is justified if the beam changes slowly compared to the timescale of plasma oscillations, and the longitudinal variation scale of the unperturbed plasma density is much longer than the plasma wavelength \cite{PPR49-229}.

\begin{figure}[h]
\includegraphics[width=82mm]{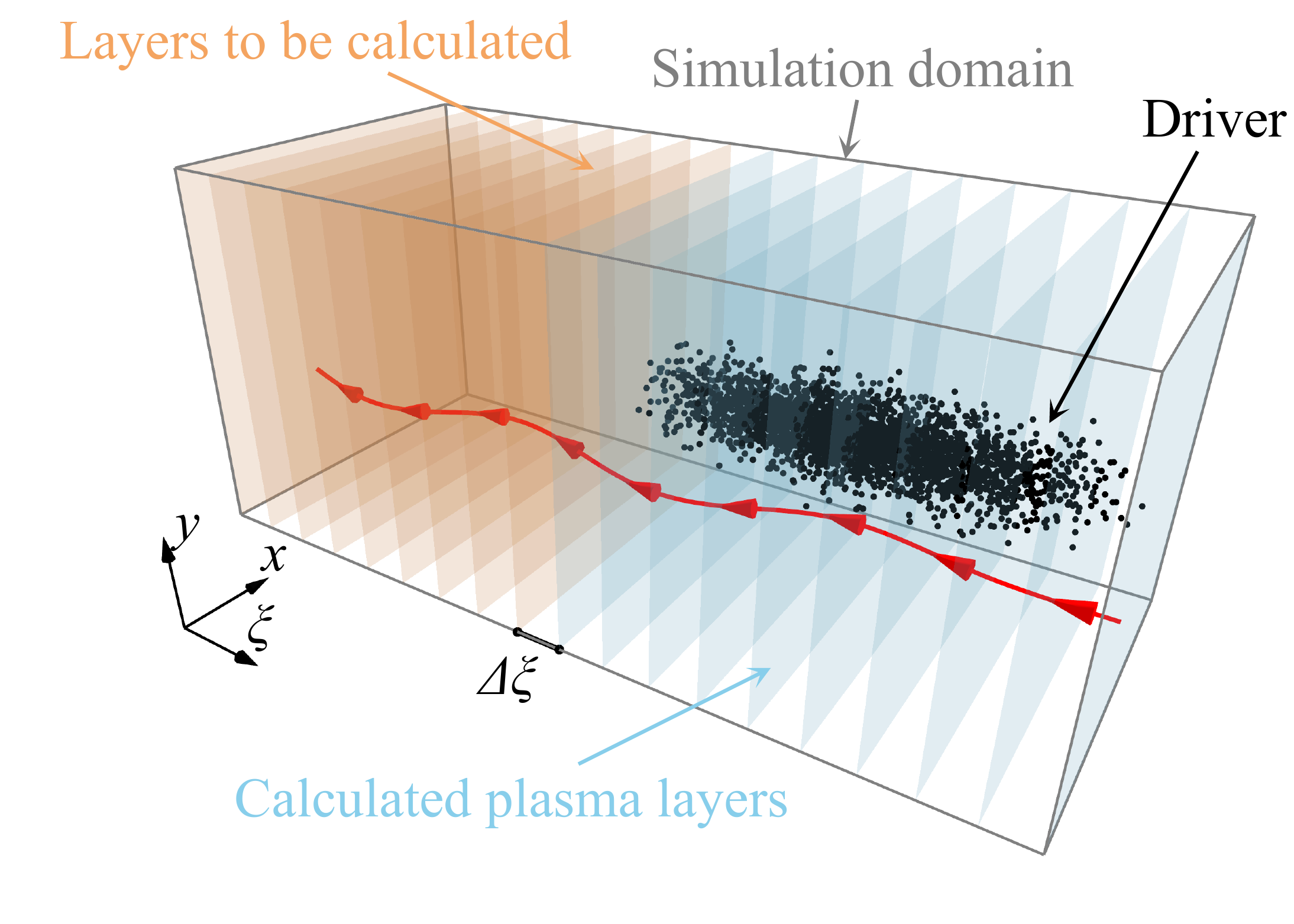}
\caption{Quasistatic simulation domain and the trajectory of a plasma ``macroparticle'' (shown in red).}\label{fig2-3d-QSA}
\end{figure}

Since plasma particles in QSA are characterized by only two spatial coordinates, their physical meaning is different from that in PIC codes. 
In PIC codes, a plasma macroparticle is a bunch of real particles. 
In the QSA model, a plasma ``macroparticle'' is an infinitely long ``string'' composed of real particles that enter the simulation domain at different times but at the same transverse coordinate and with the same initial momentum (Fig.\,\ref{fig2-3d-QSA}). 
Calculating the plasma response involves figuring out how these ``strings'' pass through the simulation domain.
Particle trajectories are built up layer by layer, starting from the front boundary, simultaneously with calculating the fields on these layers.
To stay within the familiar terminology, we will retain the term ``macroparticle'' (or simply ``particle'') to refer to ``strings'' of particles in the QSA model.

The QSA is efficient in simulating the long-term evolution of plasma wakefields at some fixed value of $z$ \cite{NatComm11-4753,PPCF64-045003}. 
Unlike the PIC model, in QSA it is possible to follow the temporal evolution of the plasma at only one value of $z$ (Fig.\,\ref{fig1-qsa}), since the behavior of nearby plasma layers is assumed to be the same but shifted in time. 
However, the duration of reliably simulated wakefield evolution may be limited by the unphysical interaction of closely spaced plasma particles.

\begin{figure}[tb]
\includegraphics{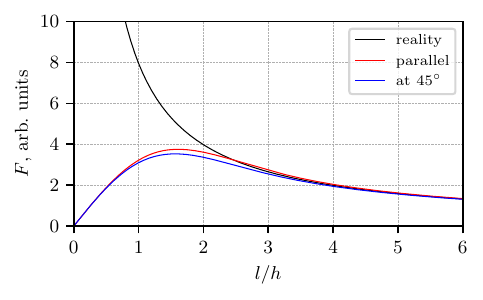}
\caption{The interaction force $F$ versus distance $l$ between two electron ``strings'' in reality and in 3d LCODE simulations with some transverse grid step $h$, if the line between electrons is parallel or at the angle of $45^\circ$ to the grid.}\label{fig3-forces}
\end{figure}

The incorrect force between closely spaced particles is a long-known effect in PIC simulations \cite{RMP55-403}, and it is also present in QSA, despite the 2d nature of the particles. 
As two particles of the same charge approach a distance shorter than the grid step, the repulsion force between them decreases, while the force between real point-like particles should tend to infinity (Fig.\,\ref{fig3-forces}).
This can be imagined as if there is an additional short-range force between the real particles.
Since wakefield behavior is determined mainly by the motion and interaction of electrons, this additional force is attractive for them and leads to the clustering of electrons, plasma heating, and eventually to incorrect results.
The recommended ways to overcome the problem (to resolve the Debye length or increase the plasma temperature \cite{H-E}) do not work for plasma wakefield acceleration, because the initial temperature of plasma electrons is very low (typically 5 eV \cite{PoP28-053104}), and it is important to keep it low to correctly simulate the long-term evolution of the wave.
The Debye length for this temperature is 300 times smaller than the minimum scale usually resolved by plasma solvers, the plasma skin depth $c/\omega_p$, where $\omega_p$ is the plasma frequency. 
Therefore, the required reduction of grid step is too drastic to be a practical solution.
Another approach is to introduce a short-range force for particles that are separated by less than some cutoff radius \cite{H-E}. 
In this case, it is necessary to generate an additional ``chaining mesh'' and linked lists, which significantly increases the numerical cost.

In this paper, we suggest a method called ``declustering'' to suppress particle clustering by modifying the particle interaction law at close distances.
Particle clustering can be seen most easily in electron density maps, where it appears as small-scale noise, so we call plasma perturbations caused by clustering ``noise perturbations''.
The declustering smooths the density profiles and greatly extends the accuracy of long-term simulations.

In Sec.\,\ref{s2:lcode}, we describe a newly developed 3d version of the quasistatic code LCODE, in which the declustering is implemented. 
In Sec.\,\ref{s4:attraction}, we show what the plasma response looks like in the presence of particle clustering. 
Then, in Sec.\,\ref{s5:noise_filter}, we detail the declustering and show how it improves the simulations.
Sec.\,\ref{s6:noise_filter2d} is devoted to an older version of declustering that was implemented in the 2d version of LCODE. Being less efficient than the new one, it nevertheless has proved very useful in simulating long plasma wakes.
Sec.\,\ref{s7:examples} presents two examples in which the declustering qualitatively changes the simulation results.
Sec.\,\ref{s8:summary} summarizes the main messages of this work.


\section{LCODE}
\label{s2:lcode}

We illustrate the problem of particle clustering and solutions to this problem with quasistatic code LCODE.
Its 2d version has been in use for almost 30 years and is well documented \cite{PoP5-785, PRST-AB6-061301,NIMA-829-350,lcode}.
Its 3d version was developed recently, so we describe its most essential physical parts (solvers) here.
Other features of the 3d code (Python implementation, parallelization, laser solver) that are not relevant to this study will be described elsewhere.

In this Section, we use dimensionless variables. 
The times are normalized to $\omega_p^{-1}$, where the plasma frequency $\omega_p = \sqrt{{4 \pi n_0 e^2}/{m}}$, and $n_0$, $e$, and $m$ are unperturbed plasma density, elementary charge, and electron mass, respectively.
The electric and magnetic fields $\vec{E}$ and $\vec{B}$ are normalized to $E_0 = m c \omega_p/e$, the current density $\vec{j}$ and charge density $\rho$ are normalized to $e n_0 c$ and $e n_0$, respectively, and the lengths are normalized to $k_p^{-1} \equiv c/\omega_p$.

\subsection{Plasma solver}
\label{s2-1:plasmalcode}

The plasma solver is the essence of the code. There is no single recipe for it, and different QSA codes use different ways of solving the equations.
The plasma solver consists of three main parts: solving equations for electromagnetic fields, moving the plasma particles, and depositing the charge and current of the particles onto the grid.

We start with Maxwell equations:
\begin{equation}\label{e_maxwell_norm}
    \operatorname{rot} \vec{E}=-\frac{\partial \vec{B}}{\partial t}, \
    \operatorname{rot} \vec{B}=\vec{j}+\frac{\partial \vec{E}}{\partial t}, \
    \operatorname{div} \vec{E}=\rho, \
    \operatorname{div} \vec{B}=0.
\end{equation}
Combining these equations and following the standard QSA (assuming that fields, charges, and currents depend on $\textbf{r}_\perp $ and $\xi$ but not on $s$) lead to six Poisson equations for the fields:
\begin{align}\label{e_poisson_1}
    \nabla^{2}_{\perp} E_{x} &= \frac{\partial \rho}{\partial x}-\frac{\partial j_{x}}{\partial \xi}, &\
    \nabla^{2}_{\perp} E_{y} &= \frac{\partial \rho}{\partial y}-\frac{\partial j_{y}}{\partial \xi}, \\
    \label{e_poisson_2}
    \nabla^{2}_{\perp} B_{x} &= \frac{\partial j_{y}}{\partial \xi}-\frac{\partial j_{z}}{\partial y}, &\
    \nabla^{2}_{\perp} B_{y} &= \frac{\partial j_{z}}{\partial x}-\frac{\partial j_{x}}{\partial \xi}, \\
    \label{e_poisson_3}
    \nabla^{2}_{\perp} E_{z} &= \frac{\partial j_{x}}{\partial x}+\frac{\partial j_{y}}{\partial y}, &\
    \nabla^{2}_{\perp} B_{z} &= \frac{\partial j_{x}}{\partial y}-\frac{\partial j_{y}}{\partial x},
\end{align}
where $\nabla^{2}_{\perp} = \partial^{2}_{x} + \partial^{2}_{y}$.

We assume perfectly conducting boundaries at $|x|= x_m$ and $|y|=x_m$. 
The plasma particles move in a narrower area $|x| < x_r$, $|y| < x_r$ and elastically reflect from its boundaries. 
This trick (as if the walls are ``painted'' with dielectric) simplifies the boundary conditions. 
The difference $x_m - x_r$ is always chosen to be larger than half the particle size, so the currents and charges turn to zero at the boundaries, and the boundary conditions are
\begin{gather}
    \label{eq:e_maxwell_border_1}
    |x|=x_m\text{:} \ B_{x}=E_{y}=E_{z}=0, \ \frac{\partial E_{x}}{\partial x}=\frac{\partial B_{y}}{\partial x}=\frac{\partial B_{z}}{\partial x}=0, \\
    \label{eq:e_maxwell_border_2}
    |y|=y_m\text{:} \ B_{y}=E_{x}=E_{z}=0, \ \frac{\partial E_{y}}{\partial y}=\frac{\partial B_{x}}{\partial y}=\frac{\partial B_{z}}{\partial y}=0.
\end{gather}
Thus, we solve the Dirichlet boundary value problem for $E_{z}$, the Neumann problem for $B_{z}$, and the mixed boundary value problems for the other field components.

To achieve numerical stability of the solver, we modify Eqs.\,(\ref{e_poisson_1})--(\ref{e_poisson_2}) by subtracting the field component with some coefficient $\mu$:
\begin{align}\label{e_helmholtz_1}
    (\nabla^{2}_{\perp} - \mu) E_{x} &= \frac{\partial \rho}{\partial x}-\frac{\partial j_{x}}{\partial \xi} - \mu \tilde E_{x}, \\
    \label{e_helmholtz_2}
    (\nabla^{2}_{\perp} - \mu) E_{y} &= \frac{\partial \rho}{\partial y}-\frac{\partial j_{y}}{\partial \xi} - \mu \tilde E_{y},\\
    \label{e_helmholtz_3}
    (\nabla^{2}_{\perp} - \mu) B_{x} &= \frac{\partial j_{y}}{\partial \xi}-\frac{\partial j_{z}}{\partial y} - \mu \tilde B_{x}, \\
    \label{e_helmholtz_4}
    (\nabla^{2}_{\perp} - \mu) B_{y} &= \frac{\partial j_{z}}{\partial x}-\frac{\partial j_{x}}{\partial \xi} - \mu \tilde B_{y}.
\end{align}
The fields marked with tildes are some predictions for the corresponding fields.
The best performance of the solver (stability and highest accuracy) is achieved for $\mu \approx 1 + h$, where $h$ is a transverse grid step.
Knowing all right-hand sides, we solve the field equations (\ref{e_poisson_3})--(\ref{e_helmholtz_4}) on the grid using Discrete Cosine and Sine Transformations (DCT and DST) of the first type.

We use the fields obtained from solving Eqs.\,(\ref{e_poisson_3})--(\ref{e_helmholtz_4}) to push the plasma and beam particles.
For the plasma particles, the dimensionless equations of motion in the QSA framework have the form
\begin{equation}\label{e_plasma_motion}
    \frac{d \mathbf{r}_{\perp}}{d \xi} = - \frac{\mathbf{v}_{\perp}}{1 - v_{z}}, \ \ \
    \frac{d \vec{p}}{d \xi} = - \frac{Q}{1 - v_{z}} \left(\vec{E}_{p} + \vec{v} \times \vec{B}_{p} \right),
\end{equation}
where $\vec{v} = \vec{p} / {\sqrt{M^2 + {\vec{p}}^2}}$, $Q$, and $M$ are the velocity, charge, and mass of the particle, respectively, $\vec{E}_{p}$ and $\vec{B}_{p}$ are the fields interpolated from the grid to the particle location.
QSA codes implement different strategies for pushing plasma particles \cite{PPCF56-084012}; we use a custom-made second-order scheme described later.

The current density $\vec{j}$ and charge density $\rho$ in the field equations are the sums of the plasma and beam (index $b$) contributions, where the plasma contribution is the sum of the electron (index $e$) and ion (index $i$) components:
\begin{equation}\label{e_current_charge_density}
    \vec{j} = \vec{j}_{b} + \vec{j}_{e} + \vec{j}_{i}, \qquad
    \rho = \rho_b + \rho_e + \rho_i.
\end{equation}
Due to their larger mass, ions move slower than electrons, and their current density is usually much lower. 
Therefore, in some cases, including those presented in this work, the ion motion can be neglected by putting $\vec{j}_{i} = 0$ and not evolving $\rho_i$.

The current and charge densities are calculated by gathering the contributions of nearby particles at each grid point using a procedure called deposition \cite{RAST9-165}.
In contrast to PIC codes, the weight of plasma particles in QSA is additionally divided by $(1 - v_{z})$ during the deposition.
Both field interpolation and deposition use a fourth-order weight function as it provides better numerical stability \cite{IPAC21-1725}.
After obtaining $\vec{j}$ and $\rho$, calculating their derivatives with respect to $x$ and $y$ is straightforward. 
The derivatives with respect to $\xi$ require currents and charges at two consecutive slices and need a special predictor-corrector loop.

The loop realizes a transition from plasma slice number $i$ to slice number $i+1$ in the direction of decreasing $\xi$. All fields and particle parameters at slice number $i$ (denoted by the corresponding superscript) are known.
The loop consists of the following sequential steps:
\begin{enumerate}

\item Propagate the plasma particles from slice $i$ to slice $i + 1$ as if there were no fields:
\begin{equation}
     \mathbf{r}^{i+1}_{\perp} = \mathbf{r}^{i}_{\perp} + \frac{\mathbf{v}^i_{\perp}}{1 - v^i_{z}} \Delta \xi,
\end{equation}
where $\Delta \xi$ is the positive longitudinal grid step.
This allows us to estimate the transverse positions of the particles at half-step $i + 1/2$,
\begin{equation}\label{e_r_halfstep}
    \mathbf{r}^{i+1/2}_{\perp} = \frac{\mathbf{r}^{i+1}_{\perp} + \mathbf{r}^{i}_{\perp}}{2},
\end{equation}    
and interpolate the fields of slice $i$ to these positions by weighted summing of several grid values \cite{H-E, RAST9-165}. We denote these fields $\vec{E}^i_p$ and $\vec{B}^i_p$.
    
\item Push the particles from slice $i$ to slice $i + 1$ according to Eqs.\,(\ref{e_plasma_motion}), using the fields obtained at the previous step. For this, we predict the particle momentum at the half-step (denoted by the superscript ``a''):
\begin{equation}
    \vec{p}^{a} = \vec{p}^i + \frac{Q}{1 - v^i_{z}} \left(\vec{E}^i_p + \vec{v}^i \times \vec{B}^i_p \right) \frac{\Delta \xi}{2},
\end{equation}
correct the momentum (denoted by the superscript ``b''):
\begin{equation}
    \vec{p}^{b} = \vec{p}^i + \frac{Q}{1 - v^{a}_{z}} \left(\vec{E}^i_p + \vec{v}^{a} \times \vec{B}^i_p \right) \frac{\Delta \xi}{2},
\end{equation}
move the plasma particle:
\begin{equation}\label{e18}
     \mathbf{r}^{i+1}_{\perp} = \mathbf{r}^{i}_{\perp} + \frac{\mathbf{v}^{b}_{\perp}}{1 - v^{b}_{z}} \Delta \xi,
\end{equation}
and, finally, calculate the particle momentum at the new slice:
\begin{equation}
    \vec{p}^{i+1} = 2 \vec{p}^b -  \vec{p}^i.
\end{equation}
    
\item Calculate the charge and current densities at slice $i + 1$, and half-sum the values at slices $i$ and $i + 1$ to get the densities at half-step $i + 1/2$.
    
\item Calculate the fields $\vec{E}^{i+1/2}$ and $\vec{B}^{i+1/2}$ from Eqs.\,(\ref{e_poisson_3})--(\ref{e_helmholtz_4}). 
The derivatives with respect to $x$ and $y$ in the right-hand sides of Eqs.\,(\ref{e_poisson_3}), (\ref{e_helmholtz_1})--(\ref{e_helmholtz_4}) are taken of $\vec{j}^{i+1/2}$ and $\rho^{i+1/2}$.
The derivatives with respect to $\xi$ use currents and charges at slices $i$ and $i+1$. The tilde-marked field predictions are taken from the known slice $i$.

\item Having newer $\mathbf{r}^{i+1}_{\perp}$ from Eq.\,(\ref{e18}), update $\mathbf{r}^{i+1/2}_{\perp}$ using Eq.\,(\ref{e_r_halfstep}). 

\item Push particles from slice $i$ to slice $i + 1$ again. This step is similar to step 2, but we use the fields $\vec{E}^{i+1/2}$ and $\vec{B}^{i+1/2}$ interpolated to the updated $\mathbf{r}^{i+1/2}_{\perp}$ instead of $\vec{E}^i_p$ and $\vec{B}^i_p$.

\item Repeat step 3.

\item Repeat step 4 using earlier obtained values of $\vec{E}^{i+1/2}$ and $\vec{B}^{i+1/2}$ as the tilde-marked field predictions. This yields updated $\vec{E}^{i+1/2}$ and $\vec{B}^{i+1/2}$.
    
\item Calculate the final fields at slice $i+1$:
\begin{equation}
    \vec{E}^{i+1} = 2 \vec{E}^{i+1/2} - \vec{E}^i, \quad
    \vec{B}^{i+1} = 2 \vec{B}^{i+1/2} - \vec{B}^i.
\end{equation}

\item Repeat steps 5--6 to obtain the final particle states at slice $i+1$.

\item Calculate the final charge and current densities at slice $i + 1$.

\end{enumerate}
After the loop is completed, we perform declustering, discussed in the following sections, using the final particle states at slice $i + 1$.

\subsection{Beam solver}
\label{s2-2:beamlcode}

The beam solver is responsible for beam initialization, beam charge and current deposition, and pushing beam particles using the fields computed by the plasma solver.

The beam can be represented differently.
One way is to directly specify $\vec{j}_b$ and $\rho_b$ and keep them unchanged.
This mode is convenient if one wants to find the plasma response to an unchanging beam and eliminate any shot noise that may appear otherwise. 
We use this mode in Sections \ref{s4:attraction}--\ref{s6:noise_filter2d} and \ref{s7-2:tr_wb}.
Another way, the beam representation as a particle ensemble, is necessary to simulate the beam evolution. 
The particles can have different or equal charges and masses and can be initially arranged in an ordered or random manner.
In Section \ref{s7-1:dispersion}, we use equal particles and randomly arrange them in both the transverse and longitudinal directions according to the given distribution.

Unlike plasma particles, beam particles have a certain longitudinal coordinate $\xi$, so they are also characterized by some longitudinal weight function.
Here we choose linear interpolation in the longitudinal direction and the same fourth-order transverse weight function that we use for plasma particles.
In contrast to the deposition in the plasma solver, there is no extra denominator $(1 - v_{z})$ in the beam particle contributions.

The equations of motion of the beam particles are similar to those used in PIC codes 
\cite{RAST9-165}:
\begin{equation}\label{e_beam_motion}
    \frac{d \mathbf{r}_{\perp}}{d s} = \mathbf{v}_{\perp}, \
    \frac{d \xi}{d s} = v_z - 1, \    
    \frac{d \vec{p}}{d s} = Q \left(\vec{E}_{p} + \vec{v} \times \vec{B}_{p}\right).
\end{equation}
We use the Higuera-Cary approach~\cite{PoP24-052104,ApJS235-21} for integrating these equations by an explicit leapfrog method. 
This approach has advantages over the other two popular methods: it preserves the phase-space volume as the Boris scheme~\cite{Boris} and preserves $\vec{E} \times \vec{B}$ drift velocity as the Vay scheme~\cite{PoP15-056701}.

By default, the integration step for Eq.\,(\ref{e_beam_motion}) equals the periodicity of field update $\Delta s$.
We need to resolve betatron oscillations of the beam particles, so the required step scales proportionally to the square root of the particle energy.
If some particles have much lower energy than others (e.g., a newly injected witness in the presence of a high-energy driver), then $\Delta s$ is determined by the lowest energy.
This slows down simulations.
In some cases, we can solve this problem with individual substepping for beam particles as follows: 
If the particle energy is less than some threshold value, we break the step $\Delta s$ into several smaller ones.
However, since individual substepping does not affect the fields, this technique only makes sense if the substepped particles have a negligible effect on the fields (e.g., there is no beam loading).

\subsection{Energy flows}
\label{s2-3:fluxlcode}

The plasma wakefield acceleration is essentially the energy transfer from the driver to the witness.
This process is most visible in the co-moving simulation window \cite{PRE69-046405} because the beams in it move slowly. 
The energy flows backward in $\xi$ from the driver, which is the energy source, to the witness, which is the energy sink.
Unused energy leaves the window through the rear boundary. 
Transverse boundaries that elastically reflect the particles do not affect the energy flux, but can do so under other reflection conditions. 
The energy flux integrated across the simulation window must be conserved between and behind the beams, and this is a convenient accuracy diagnostic of the plasma solver.

The dimensionless integral energy flux takes the form \cite{PRE69-046405,PoP25-103103, PRL130-105001}
\begin{equation}\label{ef22}
    \Psi (\xi) = \int \left( \frac{E^2 + B^2}{2} - \left[ \vec{E} \times \vec{B} \right]_z \right) dS + \Psi_p (\xi),
\end{equation}
\begin{equation}\label{ef23}
    \Psi_p (\xi) = \sum_j (\gamma_j - 1)(1 - v_{jz}) M_j,
\end{equation}
where $M_j$, $\vec{v}_j$, and $\gamma_j$ are mass, velocity, and relativistic factor of plasma particles, respectively, the integration is performed over the transverse cross-section located at some $\xi$, and the summation is over all plasma particles intersecting this cross-section.
The value of $\Psi/c$ at the corresponding $\xi$ represents either the energy released by the driver or the energy remaining behind the witness per unit length of the plasma.
The natural unit for the fluxes is
\begin{equation}
    \Psi_0 = \frac{m^2 c^5}{4 \pi e^2} \approx 2c \, \frac{\text{J}}{\text{m}}.
\end{equation}

We can also introduce the energy flux related to the kinetic energy of plasma components considered as fluids:
\begin{equation}\label{ef25}
    \Psi_f (\xi) = \int dS \sum_\alpha n_\alpha M_\alpha (\gamma_\alpha - 1)(1 - v_{\alpha z}),
\end{equation}
where $n_\alpha$, $M_\alpha$, $\vec{v}_\alpha$, and $\gamma_\alpha$ are, respectively, the density, particle mass, velocity, and relativistic factor of the plasma component $\alpha$.
In a cold plasma, the flux $\Psi_f$ equals $\Psi_p$.
However, if some plasma component has a nonzero temperature or multiple flows develop because of wave breaking or for other reasons, the difference
\begin{equation}
    \Delta \Psi = \Psi_p - \Psi_f 
\end{equation}
appears.
Thus, a nonzero $\Delta \Psi$ indicates that there are plasma particles of the same sort located at the same point, which have different velocities.

\begin{figure}[tb]
\includegraphics[width=82mm]{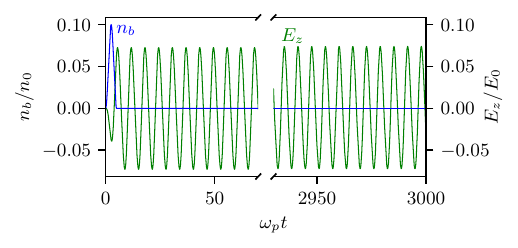}
\caption{Solution to the test problem (\ref{e2}) obtained from high-resolution 2d LCODE simulations with $\Delta \xi = h = 0.0025 k_p^{-1}$ and 10 plasma electrons per cell: electric field $E_z (t)$ and beam density $n_b (t)$ at the point $r = z = 0$.}\label{fig4-test1} 
\end{figure}

\section{Clustering of macroparticles}
\label{s4:attraction}

Let us illustrate the consequences of incorrect interparticle force by the following example \cite{NIMA-909-446}.
A short non-evolving positively charged beam of density
\begin{equation}\label{e2}
 n_b = \left\{ \begin{array}{ll}
 \displaystyle \frac{n_{b0} \, e^{-r^2/(2 \sigma_r^2)}}{2} \left[  1 - \cos \left( \frac{2 \pi \xi}{L}  \right)  \right], & -L < \xi < 0, \\
 0, & \text{otherwise},
 \end{array}\right.
\end{equation}
with
\begin{equation}\label{e3}
    n_{b0} = 0.1 n_0, \quad
    \sigma_r = k_p^{-1}, \quad 
    L = \frac{2 \sqrt{2 \pi}}{k_p}
\end{equation}
propagates in a uniform cold plasma of density $n_0$. 
The beam excites a wakefield in a nearly linear regime. 
The longitudinal electric field $E_z$ of the wave maintains its amplitude for more than 500 periods at about $0.07E_0$ (Fig.\,\ref{fig4-test1}). 
The wave period on the axis is approximately $1.00053 \cdot 2 \pi \omega_p^{-1}$.

Unless otherwise specified, we use the following parameters for 3d simulations: grid steps $h = \Delta \xi = 0.05 k_p^{-1}$, number of electrons per cell $N=4$ equals the number of immobile ions per cell. The particles are distributed regularly, the initial positions of electrons and ions coincide. 
Transverse dimensions of the simulation window are $25 k_p^{-1} \times 25 k_p^{-1}$. 
This grid is too coarse to simulate wakefield evolution over hundreds of periods \cite{IPAC13-1238} [see also Fig.\,\ref{fig7-fields}(a)], but is convenient for demonstrating the problem and its solution. 
We also present results of high-resolution simulations with $h = 0.01 k_p^{-1}$, $\Delta \xi = 0.005 k_p^{-1}$ and the same other run parameters.

\begin{figure}[tb]
\includegraphics{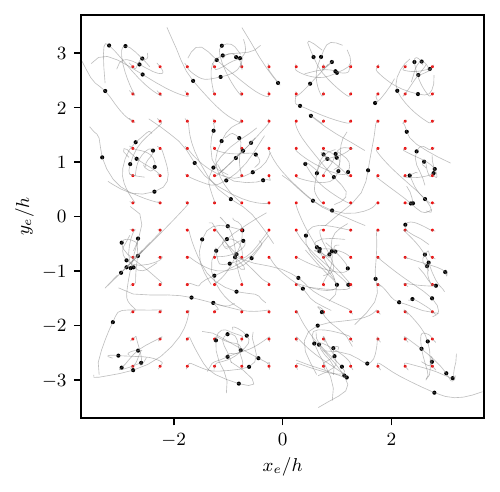}
\caption{Change in time of the relative position of individual plasma electrons. 
The distances $x_e$ and $y_e$ are measured from the center of mass of the electrons shown, thus eliminating the oscillations of these electrons as a whole in the plasma wave. 
The zero point of this figure is initially at $x=0.6 k_p^{-1}$, $y=0.5 k_p^{-1}$. 
Red dots are the initial electron locations, black dots are electron locations at $\omega_p t = 678$, and lines are electron trajectories up to time $\omega_p t = 750$.}\label{fig5-particles} 
\end{figure}

Plasma electrons, while oscillating in the plasma wave, group together into clusters spaced two grid steps apart (Fig.\,\ref{fig5-particles}). 
The characteristic clustering time depends on the wave amplitude, the higher the amplitude, the shorter the time. 
The electrons do not remain in the clusters, but continue moving relative to each other due to transverse momentum acquired during the clustering. 
The clustering process starts as an instability and at the developed stage leads to energy transfer from coherent plasma oscillations to the thermal motion of plasma electrons. 

\begin{figure}[tb]
\includegraphics{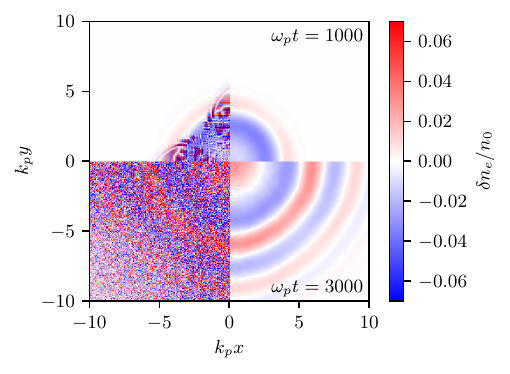}
\caption{Raw (left) and averaged (right) distributions of the electron density perturbation $\delta n_e$ at different times after beam passage.}\label{fig6-carpets}
\end{figure}
\begin{figure}[tb]
\includegraphics{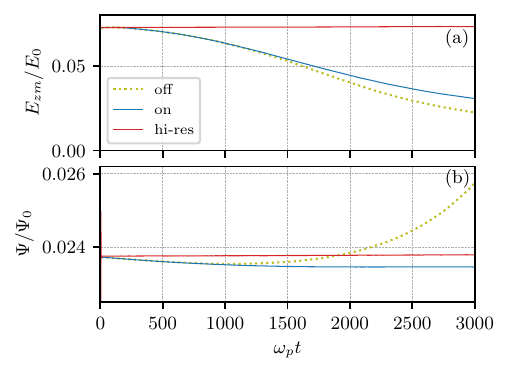}
\caption{The temporal evolution of (a) electric field amplitude $E_{zm}$ on the axis and (b) total energy flux $\Psi$ with the declustering on (solid lines) and off (dotted lines). Lines for the high-resolution simulations coincide.}\label{fig7-fields}
\end{figure}

The clusters make the plasma density distribution ``noisy'' (Fig.\,\ref{fig6-carpets}).
However, when averaged over multiple cells, the density becomes regular and smooth, so the clustering itself has no direct effect on the wakefield [Fig.\,\ref{fig7-fields}(a)], which is determined by structures of larger size (of order $k_p^{-1}$). 
The clustering also has little effect on the total plasma energy  [Fig.\,\ref{fig7-fields}(b), note the scale of energy variation]. 
The negative consequences of clustering are rapid plasma heating at the expense of wave dissipation, an increase of the effective ``size'' of macroparticles (clusters instead of individual particles), and appearance of short-range fields that can affect the emittance of accelerated beams \cite{PoP25-093112}.

\begin{figure}[tb]
\includegraphics{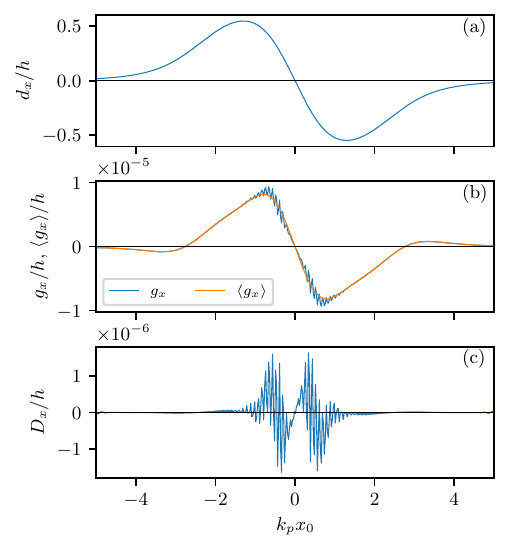}
\caption{(a) Displacement $d_x$ of plasma electrons in units of grid size $h$, (b) corresponding displacement inhomogeneity $g_x$ and its average $\langle g_x \rangle$, and (c) the noise part of the displacement $D_x$ along the line $y_0 = 0.5 k_p^{-1}$ at $\omega_p t = 300$.}\label{fig8-filtering}
\end{figure}

\begin{figure}[tb]
\includegraphics{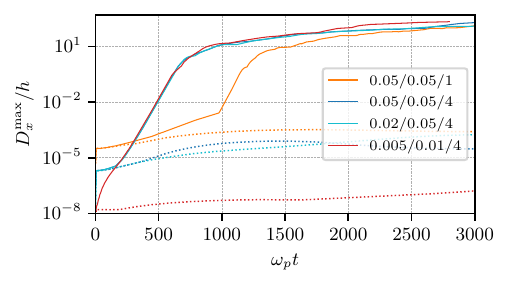}
\caption{Temporal growth of the maximum noise displacements $D_x^\text{max}$ at different grid steps with the declustering on (dotted lines) and off (solid lines). The legend shows run parameters in format $k_p \Delta \xi / k_p h / N$.}\label{fig9-logline}
\end{figure}

To fix the problem, it is necessary to detect clustering as early as possible.
For this purpose, we consider displacements of particles relative to their initial positions $\textbf{r}_{0\perp}$:
\begin{equation}
    \textbf{d} = \textbf{r}_{\perp} - \textbf{r}_{0\perp}.
\end{equation}
The displacement is usually dominated by the particle motion in the plasma wave [Fig.\,\ref{fig8-filtering}(a)]. 
Next, we introduce displacement inhomogeneities with respect to neighboring particles:
\begin{equation}\label{e5}
    \textbf{g}_{i,j} = \textbf{d}_{i,j} 
    - \frac{\textbf{d}_{i+1,j} + \textbf{d}_{i-1,j} + \textbf{d}_{i,j+1} + \textbf{d}_{i,j-1}}{4},
\end{equation}
where the subscripts $i,j$ denote the ordinal numbers of particles in the directions $x$ and $y$; we write these subscripts only when the formulas contain quantities referring to different particles. 
The displacement inhomogeneities are much smaller than the displacements themselves, but they are also often dominated by the plasma wave [Fig.\,\ref{fig8-filtering}(b)]. 
To extract the noise component of the displacement, we average $\textbf{g}$ over some area covering several particles and subtract this average:
\begin{equation}\label{e31}
    \textbf{D} = \textbf{g} - \langle \textbf{g} \rangle,
\end{equation}
where angle brackets denote averaging. 
This procedure allows us to detect the noise at the earliest stages of its growth [Fig.\,\ref{fig8-filtering}(c)], typically at amplitudes less than $10^{-4} h$, as can be seen from the time dependence of $D_x^\text{max}$, the maximum of $D_x$ over all particles (Fig.\,\ref{fig9-logline}).

\begin{figure}[tb]
\includegraphics{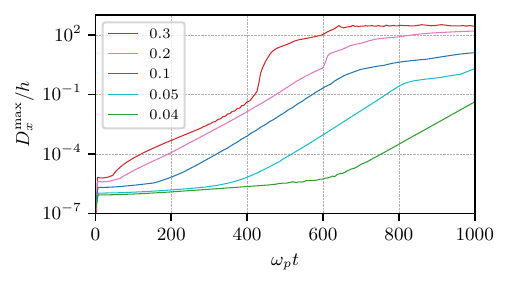}
\caption{Temporal growth of the maximum noise displacements $D_x^\text{max}$ for different wave amplitudes controlled by varying the driver density $n_{b0}$ in formula (\ref{e2}). 
The legend shows the ratio $n_{b0}/n_0$. 
Faster displacement growth for large $n_{b0}/n_0$ is due to wavebreaking.}\label{fig10-logline2}
\end{figure}

If no special measures are taken, the noise displacements grow exponentially. 
The growth rate is independent of grid size and number of electrons per cell $N$ (Fig.\,\ref{fig9-logline}) and weakly dependent on the wave amplitude (Fig.\,\ref{fig10-logline2}). 
However, the larger the wave amplitude, the higher the initial level of the instability, and the faster the noise grows to a high level.

\section{Declustering}
\label{s5:noise_filter}

Having determined the noise displacement of plasma particles, we can influence it by modifying the equations of particle motion. 
We add an extra transverse momentum to each plasma electron at each step in $\xi$:
\begin{multline}\label{e7}
    \Delta \textbf{p}_{i,j} = - K \Delta \xi \Biggl[
    F_{i,j} \textbf{D}_{i,j} 
    - \frac{F_{i+1,j} \textbf{D}_{i+1,j} + F_{i-1,j} \textbf{D}_{i-1,j}}{4} \\
    - \frac{F_{i,j+1} \textbf{D}_{i,j+1} + F_{i,j-1} \textbf{D}_{i,j-1}}{4} \Biggr] \\
    - \kappa \Delta \xi F_{i,j} \left[ 
    \textbf{D}_{i,j} - \textbf{D}_{i,j}^\text{prev} \right],
\end{multline}
where 
\begin{equation}
    F_{i,j} = \left\{ \begin{array}{lcl}
        \displaystyle 1 - |\textbf{D}_{i,j}|^2 / D_0^2, & \quad & |\textbf{D}_{i,j}| < D_0, \\
    0, && \text{otherwise},
 \end{array}\right.
\end{equation}
$\textbf{D}_{i,j}^\text{prev}$ is the noise displacement at the previous step in $\xi$, and $K$, $\kappa$, and $D_0$ are adjustable parameters. 

The coefficient $K$ characterizes the restoring force. 
If a particle is displaced in some direction relative to where it should be in large-scale perturbations, this force pushes it back. 
We assume that the restoring force is exerted by nearby particles, so a counterforce must act on them to fulfill Newton's third law. The fractional terms in Eq.\,(\ref{e7}) are responsible for the counterforce.

The restoring force alone cannot suppress the noise. 
Since it is non-dissipative, it forces noise perturbations to propagate as a wave with some group velocity. 
Therefore, we introduce a damping term (second term) into Eq.\,(\ref{e7}) and control the damping with the coefficient $\kappa$.

The third coefficient $D_0$ limits the declustering range. 
It smoothly reduces the force to zero if the displacement is large enough. 
The motion of plasma particles in the wake can be accompanied by strong inhomogeneities in interparticle distances, for example, at wave breaking. 
The declustering should have minimal effect on these perturbations. 
Real and non-physical inhomogeneities can be distinguished by the growth law. 
The physical (or real) inhomogeneities, if any, quickly grow to large values regardless of declustering, while the clustering stays at very low level for some time and can be filtered out. 
To get a sense of the scales, let us estimate the displacements caused by a single plasma electron if its charge is not balanced by an ion. 
If there are $N$ electrons per cell, then the linear charge density of one (macro)electron is 
\begin{equation}
    \lambda = \frac{e n_0 h^2}{N}.
\end{equation}
This ``electron string'' pushes nearby plasma electrons in different directions by a transverse electric field of the scale 
\begin{equation}
    E_\lambda \sim \frac{2 \lambda}{h} \sim \frac{2 e n_0 h}{N},
\end{equation}
causing oscillations with plasma frequency $\omega_p$ and amplitude 
\begin{equation}
    d_\lambda \sim \frac{e E_\lambda}{m \omega_p^2} \sim \frac{h}{2 \pi N}.
\end{equation}
These oscillations driven by individual electrons can be seen on electron density maps in some regimes (see Fig.\,\ref{fig18-visualization}).
Displacement inhomogeneities caused by a single wandering macroparticle are of the same order: 
\begin{equation}
    D_\lambda / h \sim \frac{1}{2 \pi N} \gtrsim 10^{-2}.
\end{equation}
This displacement is large enough that we can choose a value of $D_0 \ll D_\lambda$ such that it is 1-2 orders of magnitude greater than the noise detection level ($D_0 \sim 10^{-3} h$, Fig.\,\ref{fig9-logline}). In this way, we can suppress the noise without affecting the physical processes in the wave.

\begin{figure}[tb]
\includegraphics{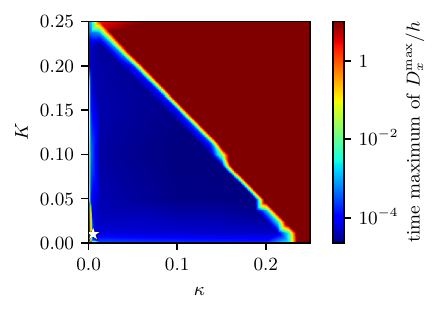}
\caption{Stability region on the plane of declustering parameters $\kappa$ and $K$, represented as the maximum of $D_x^\text{max}$ over the simulation time up to $3000 \omega_p^{-1}$. The white star shows the working point ($K=0.01$, $\kappa=0.005$) used in the illustrative examples of Sec.\,\ref{s4:attraction}. }\label{fig11-triangle}
\end{figure}

The optimal values of coefficients $K$ and $\kappa$ depend on the problem being solved. 
The coefficients must obey
\begin{equation}\label{e9}
    K + \kappa < \frac{h^2}{\Delta \xi^2 N}.
\end{equation}
Otherwise, neighboring particles start to oscillate in antiphase at the Nyquist frequency with exponentially growing amplitude, and the declustering becomes unstable (Fig.\,\ref{fig11-triangle}).
Typically, the optimal values are the smallest values sufficient to suppress clustering.
Experience is gradually being gained to automatically select the coefficients in the code as best as possible.

\section{Declustering in two-dimensional geometry}
\label{s6:noise_filter2d}

The declustering implemented in 2d version of LCODE is less efficient than the one described above because it damps only one instability mode. However, it allowed simulations of long plasma wakes (more than 100 periods \cite{NIMA-909-446, PPCF62-125023, PPCF62-115025, PPCF63-125027}),  so it deserves to be described. The formulas given below refer to the axisymmetric case, their modification for the plane case is straightforward.

Suppose that the plasma electron density $n_e$ has a short-scale modulation
\begin{equation}\label{2e1}
    \delta n = - A_j \cos (k_h \delta r), \qquad k_h = \pi/h,
\end{equation}
where $\delta r$ is the distance to the grid node with number $j$. We can find the modulation amplitude $A_j$ in the vicinity of this node from the density in neighboring nodes:
\begin{equation}\label{2e2}
    A_j \approx (n_{e,j+1} + n_{e,j-1} - 2n_{e,j})/4.
\end{equation}
There should appear a radial electric field $E_r$, which is determined by the Poisson equation
\begin{equation}\label{2e3}
    \frac{1}{r} \frac{\partial}{\partial r} r E_r = 4 \pi \left( e n_i - e n_e + \rho_b \right) - \frac{\partial E_z}{\partial \xi},
\end{equation}
directed so as to reduce the density perturbations.
Suppose that the longitudinal electric field $E_z$, the beam charge density $\rho_b$, and the ion density $n_i$ have no short-scale modulation. Then the short-scale component of the field, $\delta E_r$, is determined only by the modulation of the electron density:
\begin{equation}\label{2e4}
    \frac{1}{r} \frac{\partial}{\partial r} r \delta E_r =  4 \pi e A_j \cos (k_h \delta r).
\end{equation}
Let us solve this equation using complex amplitudes:
\begin{equation}\label{2e5}
    \delta n = \text{Re} \, (A_j e^{i k_h \delta r}), \qquad \delta E_r = \text{Re} \, (B_j e^{i k_h \delta r}).
\end{equation}
\begin{equation}\label{2e6a}
    \frac{B_j}{r} + i k_h B_j = 4 \pi e A_j,
\end{equation}
\begin{equation}\label{2e6}
    B_j = \frac{4 \pi e A_j}{i k_h + 1/r} = \frac{4 \pi e A_j h}{i \pi + 1/j} = \frac{4 \pi e A_j h}{\pi^2 + 1/j^2} \left( \frac{1}{j} - i \pi \right).
\end{equation}
Returning to the real values, we obtain
\begin{equation}\label{2e7}
    \delta E_r = \frac{4 \pi^2 e A_j h}{\pi^2 + 1/j^2} \sin (k_h \delta r) + \frac{4 \pi e A_j h}{j \pi^2 + 1/j} \cos (k_h \delta r).
\end{equation}
The force $(-e \delta E_r)$ caused by this short-scale component of the field is added to other forces acting on individual electrons at each step in $\xi$. The value of $A_j$ is taken at the grid node closest to the electron.

\begin{figure}[tb]
\includegraphics{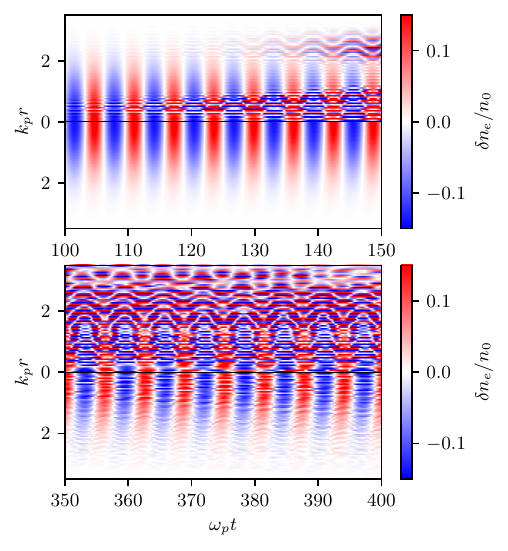}
\caption{Plasma response to the beam (\ref{e2}) calculated with the axisymmetric code: Fragments of electron density maps with the declustering off (upper halves) and on (lower halves) at different delays after beam passage. }\label{fig12-2d}
\end{figure}

The above algorithm delays noise growth and associated unphysical effects by suppressing the fastest growing mode of the instability (Fig.\,\ref{fig12-2d}). However, it cannot completely remove the noise because perturbations with smaller wave numbers continue to grow.

\section{Examples when declustering is necessary}
\label{s7:examples}

\subsection{Beam self-modulation}
\label{s7-1:dispersion}

This example shows how the clustering adds a chaotic component to the transverse wakefield.
The effect disappears with declustering turned on.

\begin{figure}[tb]
\includegraphics{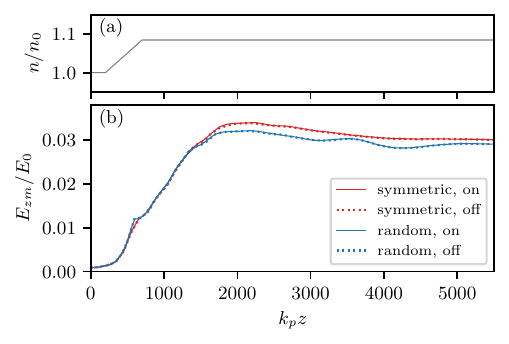}
\caption{Self-modulation test problem: (a) plasma density $n$ and (b) maximum excited wakefield $E_{zm}$ versus propagation distance $z$. The legend indicates whether the beam particles are initially distributed symmetrically or randomly in space, and whether the declustering is on or off.} \label{fig13-smisim1}
\end{figure}

\begin{figure}[tb]
\includegraphics{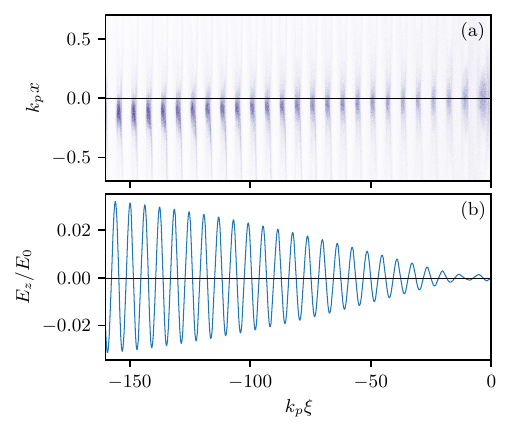}
\caption{Self-modulation test problem: (a) Beam portrait (beam density $n_b(x,y,\xi)$ integrated over coordinate $y$) at $z=2000 k_p^{-1}$ for the case of random initial distribution of beam particles and the declustering enabled; (b) the on-axis electric field $E_z$ excited by this beam.}\label{fig14-smisim2}
\end{figure}

Consider the test problem used in Ref.~\cite{PoP22-103110} to illustrate self-modulation of a long particle beam. A positron beam initially (at $z=0$) has a relativistic factor $\gamma_b = 1000$, an angular spread of $2 \times 10^{-4}$, and a density
\begin{equation}\label{e22}
 n_b = \left\{ \begin{array}{ll}
 \displaystyle n_{b0} \, e^{-r^2/(2 \sigma_r^2)}, & \xi < 0, \\
 0, & \xi \geq 0,
 \end{array}\right.
\end{equation}
with
\begin{equation}\label{e23}
    n_{b0} = 4 \times 10^{-3} n_0, \qquad
    \sigma_r = 0.5 k_p^{-1}.
\end{equation}
The beam propagates in a cold, radially uniform plasma whose density $n$ depends on $z$ [Fig.\,\ref{fig13-smisim1}(a)]:
\begin{equation}
    \frac{n}{n_0} = \left\{ \begin{array}{ll}
 \displaystyle 1, & k_p z \leq 200, \\
 \displaystyle 1 + 1.7 \times 10^{-4} (k_p z - 200), & 200 < k_p z < 700, \\
 1.085, & k_p z \geq 700.
 \end{array}\right.
\end{equation}
The beam undergoes self-modulation initiated by the steep leading edge, splits into short microbunches spaced approximately one plasma period apart (Fig.\,\ref{fig14-smisim2}), and continues to propagate in a bunched state with no field degradation  [Fig.\,\ref{fig13-smisim1}(b)] due to the increase in plasma density experienced during the self-modulation.

In simulations of this test problem, $h = \Delta \xi = 0.02 k_p^{-1}$, $\Delta s = 50 k_p^{-1}$, $N=1$, transverse dimensions of the simulation window are $5 k_p^{-1} \times 5 k_p^{-1}$, and $N_b = 1.6 \times 10^7$ macroparticles are used for the beam. In variants with declustering, $K=0.003$, $\kappa=0.001$, and deactivation of declustering is disabled ($D_0$ is large) because no wavebreaking is expected.

\begin{figure}[tb]
\includegraphics{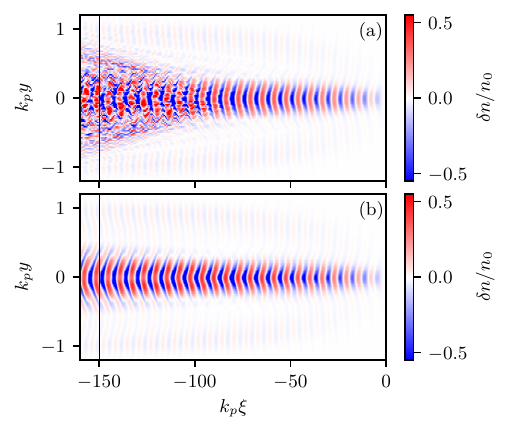}
\caption{Perturbations of the plasma electron density $\delta n$ caused by a self-modulating beam (\ref{e22}) at $x=0$, $z=1000 k_p^{-1}$ with the declustering off (a) and on (b).
The beam macroparticles are initially randomly distributed in space. 
The thin vertical lines mark the cross-section where the $E_y$ field is shown in Fig.\,\ref{fig16-Ex}.}\label{fig15-dispersion}
\end{figure}
\begin{figure}[tb]
\includegraphics{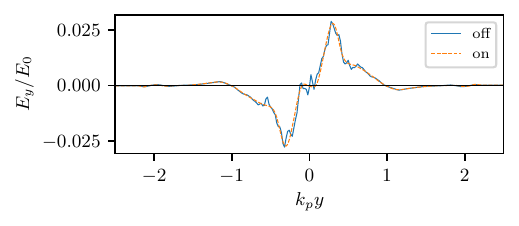}
\caption{The transverse electric field $E_y$ along the line $x=0$, $\xi=-150 k_p^{-1}$ at $z = 1000 k_p^{-1}$  with the declustering off and on.}\label{fig16-Ex}
\end{figure}

Without the declustering, small-scale density perturbations occur in the plasma and the plasma electrons heat up.
The perturbations have a non-zero group velocity and propagate transversely beyond the beam area [Fig.\,\ref{fig15-dispersion}(a)]. 
The clustering does not affect the wakefield amplitude [the solid and dashed lines in Fig.\,\ref{fig13-smisim1}(b) coincide] but introduces a chaotic perturbation to the transverse wakefield (Fig.\,\ref{fig16-Ex}). 
Perturbations of this kind can lead to unphysical growth of the accelerated beam emittance \cite{PoP25-093112} and must be avoided in accurate simulations. 
With the declustering, the density profile is smooth, the density perturbations remain within the beam area [Fig.\,\ref{fig15-dispersion}(b)], and the transverse electric field is also smooth (Fig.\,\ref{fig16-Ex}).

\begin{figure}[tb]
\includegraphics{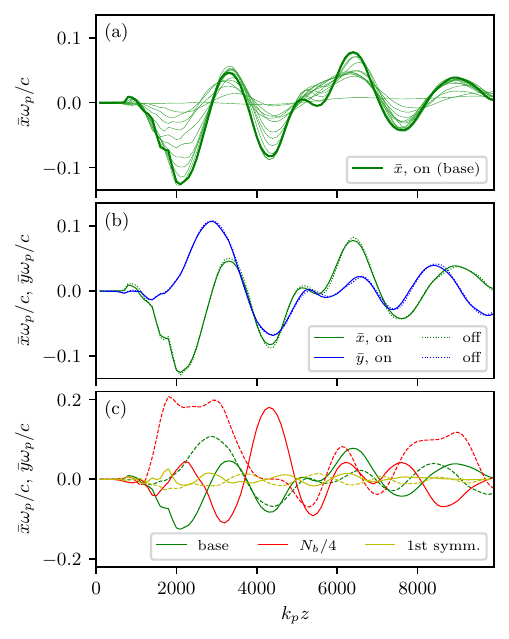}
\caption{Deviation of microbunches from the axis: 
(a) deviations $\bar x$ of each odd microbunch (thin lines) and of the last (25th) microbunch (thick line) for random initial distribution of beam particles and declustering enabled, 
(b) deviations $\bar x$ and $\bar y$ of the last microbunch with the declustering on (solid lines) and off (dotted lines), 
(c) deviations $\bar x$ (solid lines) and $\bar y$ (dashed lines) of the last microbunch for variants with $N_b$ randomly distributed macroparticles in the beam, with reduced ($N_b/4$) number of beam macroparticles, and with symmetric distribution of macroparticles in the first microbunch.}\label{fig17-hosing}
\end{figure}

Clean simulations of self-modulation provide insight into the cause of slow long-wavelength transverse oscillations of microbunches.
These oscillations are sometimes evident in simulations \cite{PF}, and even visible in experiments \cite{Huther}. 
They are also present in our simulations: Fig.\,\ref{fig14-smisim2}(a) shows the microbunches at the moment of almost maximum deflection.
To visualize the oscillations better, we plot the average transverse coordinates of individual microbunches, $\bar x$ and $\bar y$, as functions of distance traveled (Fig.\,\ref{fig17-hosing}). 
Only beam particles with radial position $|\mathbf{r}_\perp| < k_p^{-1}$ are taken into account when calculating the average coordinates to avoid the influence of defocused particles.
In some variants, beam particles initially have a symmetric distribution either in the whole beam or only in the first microbunch (at $k_p \xi> -2 \pi$); this means that each beam particle has three counterparts mirrored with respect to the planes $(x,\xi)$, $(y,\xi)$, and the axis $\xi$.

Several observations follow from the obtained dependencies. 
First, transverse deviations increase along the bunch train, but reach saturation at some point. 
This is seen in Fig.\,\ref{fig17-hosing}(a), where the deviations at $k_p z \sim 2000$ increase monotonically over the first 20 microbunches and then saturate [Fig.\,\ref{fig14-smisim2}(a)]. 
Therefore, long-wavelength hosing is an instability in the sense that it amplifies the deviations along the bunch train: later microbunches are displaced in the same direction as earlier microbunches, but to a greater extent. 

Second, the growth time of the instability and the period of transverse oscillations are comparable to the inverse betatron frequency
\begin{equation}
    \tau_b = \sqrt{2 \gamma_b/n_{b0}} \sim 700 \omega_p^{-1}
\end{equation}
of the original (unmodulated) beam and, hence, to the growth times of the usual self-modulation \cite{PRL107-145003, PRL107-145002} or hosing \cite{PoP4-1154, PRE86-026402}. 
The deviations reach a maximum at the time of full bunching (in our case at $k_p z \sim 2000$). 
Therefore, the leading part of the beam plays a dominant role in long-wavelength hosing, since the trailing part experiences a much stronger focusing force (when the wave is resonantly driven by many microbunches) and has much shorter characteristic times of transverse dynamics.

Third, due to stronger focusing, the trailing microbunches quickly get out of resonance with the driving force oscillating at the frequency of the leading microbunches and copy their deviation without amplification of the amplitude [Fig.\,\ref{fig14-smisim2}(a)]. 
Therefore, the hosing amplitude after the first maxima does not grow further with propagation distance and even slowly decreases (Fig.\,\ref{fig17-hosing}, all variants).

Fourth, the long-wavelength hosing does not depend on declustering and therefore cannot be a consequence of incorrectly simulated plasma response, as follows from the coincidence of the solid and dotted lines in Fig.\,\ref{fig17-hosing}(b).

Fifth, the hosing amplitude increases with decreasing number of beam macroparticles used in simulations [Fig.\,\ref{fig17-hosing}(c)]. 
If the particles in the first microbunch are distributed symmetrically, the amplitude of oscillations is much lower. 
If the particles are distributed symmetrically in the whole beam, no hosing develops, the microbunches propagate strictly along the axis, and the excited wakefield is slightly larger than for random distribution and hosing [Fig.\,\ref{fig13-smisim1}(b)].

Therefore, the long-wavelength hosing is a consequence of asymmetry in the front part of the beam. 
This asymmetry arises in simulations due to the random positioning of beam macroparticles. 
In real experiments, the long-wavelength hosing may be weaker because of much larger number of particles in the beam, or stronger if the beam has initial asymmetry.

\begin{figure}[tb]
\includegraphics{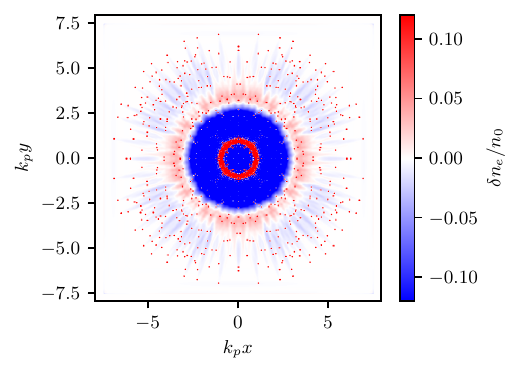}
\caption{Electron density perturbation $\delta n_e$ at $\omega_p t = 78$ (soon after the wavebreaking) in 3d simulations with $k_p h = 0.03$. Red dots are individual electrons (macroparticles), weakly colored tails behind them are their wakefields.}\label{fig18-visualization}
\end{figure}

\subsection{Transverse wavebreaking}
\label{s7-2:tr_wb}

Transverse wavebreaking is an event in which the trajectories of plasma electrons, initially at different radii, intersect, causing one or both electrons to drop out of collective motion in the wave (Fig.\,\ref{fig18-visualization}).
Transverse wavebreaking plays an important role in the plasma wakefield acceleration because it can lead to electron trapping and acceleration by the plasma wave \cite{PRL78-4205}, wave dissipation \cite{PoP25-103103}, appearance of radially ejected electrons \cite{PRL112-194801,PoP23-103112,PPCF63-055002,PoP29-023104} and even field growth \cite{imotion2}.

\begin{figure}[tb]
\includegraphics{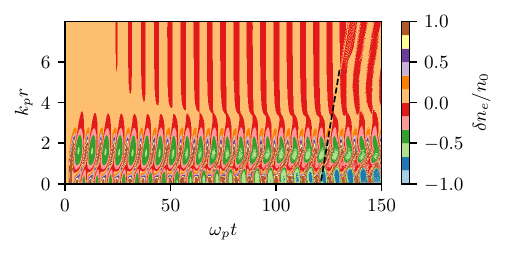}
\caption{Wavebreaking in the reference 2d run: electron density map with trajectories of radially escaping electrons highlighted and extrapolated by a dashed line.}\label{fig19-reference}
\end{figure}

As expected, transverse wavebreaking is sensitive to plasma electron clustering.
Let us illustrate this with the example (\ref{e2}) but with $n_{b0} = 2 n_0$.
In the selected variant, when the wave is driven by a short bunch, the wavebreaking occurs solely due to wave nonlinearity.
There is no ion motion \cite{PRL86-3332,PoP10-1124,PRL109-145005,PoP21-056705,PoP25-103103} or long beam \cite{PoP25-063108, PoP29-023104} that could enhance wavefront distortion, so the wavebreaking position is particularly sensitive to simulation conditions.
In 3d simulations of this section, $h = \Delta \xi$, there is 1 electron and 1 immobile ions per cell, transverse dimensions of the simulation window are $16 k_p^{-1} \times 16 k_p^{-1}$, $K=0.3$, $\kappa = 0.1$, and $D_0 = 10^{-3}$.
As a reference to compare with, we take a high-resolution 2d run with $h = \Delta \xi = 0.0025 k_p^{-1}$, 10 electrons per cell, and declustering turned on (Fig.\,\ref{fig19-reference}). 

\begin{figure}[tb]
\includegraphics{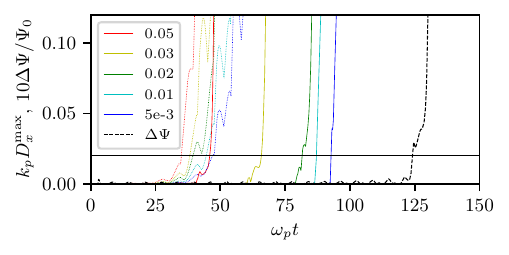}
\caption{Maximum noise displacement $D_x^\text{max}(t)$ in 3d simulations with different resolutions and with the declustering on (solid lines) and off (dotted lines). Grid steps $h$ and $\Delta \xi$ in units of $k_p^{-1}$ are given in legend. 
The black dashed line shows the nonhydrodinamic part of the energy flux $\Delta \Psi (t)$ in the reference 2d run. 
The first crossings of the horizontal black line are considered the moments of wavebreaking.}
\label{fig20-Dmax} 
\end{figure}

The question arises, however, how to determine the precise moment of wavebreaking. 
Intersection of trajectories is an impractical criterion because even in the axisymmetric geometry, trajectories can intersect due to the noise we are studying. 
In 3d geometry, it is not at all clear which trajectories to compare, since the electrons are not ordered by radial position, but distributed over the $(x,y)$ plane. 
In the reference 2d run, we find the wavebreaking point either by extrapolating the trajectories of radially runaway electrons (Fig.\,\ref{fig19-reference}) or by measuring the non-hydrodynamic contribution to the energy flux $\Delta \Psi$ \cite{PoP25-103103} and requiring $\Delta \Psi > 0.002 \Psi_0$ (Fig.\,\ref{fig20-Dmax}); both methods give the same value. 
In 3d runs, we use the criterion $D_x^\text{max} > 0.02 k_p^{-1}$. 
When the wave breaks, the displacement inhomogeneity (\ref{e31}) of some particles increases abruptly (Fig.\,\ref{fig20-Dmax}), which makes it possible to localize the wavebreaking in time and space.

\begin{figure}[tb]
\includegraphics{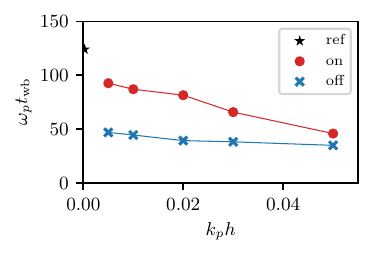}
\caption{Dependence of the wavebreaking time $t_\text{wb}$ on the grid step $h$ with the declustering on and off. The star shows the wavebreaking time in the reference 2d run.}\label{fig21-converge}
\end{figure}

The wavebreaking moment $t_\text{wb}$ observed in simulations depends on the grid steps $h$ and $\Delta \xi$ (Figs.\,\ref{fig20-Dmax}, \ref{fig21-converge}). 
The wave breaks earlier with larger steps. 
As the grid step decreases, the wavebreaking time approaches that observed in the reference run, but only when the declustering is enabled. 
If not, then the noise provokes earlier wavebreaking for any resolution.
Therefore, simulations performed without declustering can potentially overestimate many important physical effects associated with wavebreaking.

\section{Summary}
\label{s8:summary}

Simulations of plasma wakefield acceleration are challenging for many reasons, one of which is that the wave scale ($\sim c/\omega_p$) is much larger than the Debye length.
Attempting to resolve both would make simulations prohibitively expensive. 
Therefore, grid steps far exceeding the Debye length are used, which compromises the accuracy of PIC simulations.
While this inconsistency does not notably impact short-term processes spanning several wave periods, long-term simulations lead to clustering and unphysical heating of plasma electrons.

The latter undesirable effects can be avoided by slightly modifying the law of plasma particle motion, which is implemented in the newly developed 3d version of LCODE.
The new feature enables much cleaner simulations of long-term wave evolution.
In particular, when the numerical heating of plasma electrons is suppressed, the wake chaotization and premature transverse wavebreaking of a moderately nonlinear plasma wave disappear.

\acknowledgements

This work was supported by the Russian Science Foundation, Project No. 23-12-00028.

\end{document}